\documentclass[9pt,twocolumn,twoside]{pnas-new}

\usepackage{amsmath}
\usepackage[version=4]{mhchem}
\usepackage{siunitx}

\templatetype{pnasresearcharticle} 

\renewcommand{\bibliography}[1]{}

\begin{document}

\title{Deuterated water and the formation of the
satellites of Uranus}

\author[a,1]{Michael E. Brown}
\author[a]{Matthew Belyakov}
\author[a]{Swaroop Chandra}
\author[a]{M. Ryleigh Davis}
\author[a]{Merritt McDowell}
\author[a]{Ashma Pandya}
\author[a]{Kevin T. Trinh}
\author[b]{Samantha K. Trumbo}

\affil[a]{Division of Geological and Planetary Sciences, California Institute of Technology, Pasadena, CA 91125}
\affil[b]{Department of Astronomy \& Astrophysics, University of California, San Diego, La Jolla, CA 92093, USA}

\leadauthor{Brown}

\significancestatement{
While the nearly perpendicular tilt of Uranus’s spin axis could have been caused by a giant impact, the origin of the satellites’ similarly tilted orbits remains unclear. Current hypotheses include formation from debris produced by the tilting impact, destruction and reaccumulation of a pre‑existing satellite system, or accretion of passing material from the outer solar system. Using measurements from the James Webb Space Telescope, we show that the deuterium‑to‑hydrogen ratio in the satellites’ water ice is nearly five times higher than that of Uranus, ruling out scenarios in which substantial amounts of Uranian material were incorporated into the satellites. The marginally higher deuterium abundance on the innermost satellite, Miranda, may offer a crucial clue to the satellite's origins.
}
\authorcontributions{MEB conceived the project,
analyzed the data, and
wrote the manuscript. MB and MRD obtained and reduced the Miranda
data. CS, MEB, and AP obtained and analyzed the laboratory data. All authors discussed results and
implications and provided feedback on the manuscript.}

\authordeclaration{The authors declare that they have
no competing interests.}

\correspondingauthor{\textsuperscript{1}To whom correspondence should be addressed. E-mail: mbrown@caltech.edu}


\begin{abstract}
The satellites of Uranus orbit in a low‑eccentricity, equatorial plane that is tilted by 98$^\circ$ relative to the solar system—a geometry that mirrors Uranus’s extreme axial tilt. Although a giant impact could have tipped Uranus, how the satellites came to share this orientation remains uncertain. Proposed formation pathways include primordial accretion followed by reorientation, formation from debris generated by the tilting impact, and reaccretion from a massive ring produced by the tidal disruption of passing bodies from the outer solar system. Current observations do not discriminate among these scenarios.
Using the James Webb Space Telescope, we measured the deuterium‑to‑hydrogen (D/H) ratio in the water ice of the five regular satellites of Uranus. We find an average D/H ratio of 
$2.1 \pm 0.2 \times 10^{-4}$, nearly five times higher than that of Uranus and comparable to values measured in comets. This enrichment is inconsistent with any formation scenario in which substantial Uranian material was incorporated into the satellites, thereby excluding models that require significant mixing in an impact‑derived vapor disk.
The observed D/H ratios are instead compatible with models in which the satellites accreted from material that remained largely separate from Uranus, such as debris from a disrupted pre‑existing satellite system or from a tidally captured outer solar system body. The innermost regular satellite, Miranda, exhibits a marginally elevated D/H ratio (2.8$\sigma$ above the average of the outer satellites), potentially indicating a distinct formation history or source of water and offering an important clue for distinguishing among competing models.

\end{abstract}

\doi{\url{www.pnas.org/cgi/doi/10.1073/pnas.2519276123}}

\maketitle
\thispagestyle{firststyle}
\ifthenelse{\boolean{shortarticle}}{\ifthenelse{\boolean{singlecolumn}}{\abscontentformatted}{\abscontent}}{}


\dropcap{A}ll giant planets in our solar system
have satellites, but each system has unique properties
suggesting individual formation pathways. The four
large Galilean satellites of Jupiter seem the
most straightforward to explain, as they appear
to have formed in a circumplanetary disk in
some ways analogous to the formation of the planets
in the circumsolar disk \cite{lunine_formation_1982,
Peale_origin_1999, Canup_formation_2002, Canup_common_2006}. 
Saturn, with its rings, series of mid-sized moons, and single 
large satellite is more complicated, but a combination
of satellite formation, destruction, re-accretion from
the rings, and fast tidal migration to their present
positions appears a promising scenario \cite{Canup_origin_2010,Charnoz_accretion_2011,asphaug_late_2013,blanc_understanding_2025}.
Neptune is the only giant planet with no regular 
satellite system, instead hosting the massive Triton
which appears to be a captured body from the outer
solar system \cite{goldreich_neptunes_1989,banfield_dynamical_1992,agnor_neptunes_2006,rufu_tritons_2017}. Of the giant
planet satellite systems, the origin of that orbiting
Uranus remains the least understood. 

Uranus has a series of 5 regular satellites, ranging in diameter
from nearly 470 km (the inner satellite Miranda) to almost
1700 km (the outer pair Titania and Oberon), { with a total mass
approximately 10$^{-4}$ times that of Uranus}. The central pair, Ariel
and Umbriel, are about 1160 km in diameter. 
These regular satellites
all orbit in a prograde sense in
the equatorial plane of Uranus, but the planet itself
has an obliquity of nearly 98 degrees.
The
source of this unexpected tilt is unknown.
An ancient giant impact is often taken to be
the cause \cite{safronov_sizes_1966, morbidelli_explaining_2012, ida_uranian_2020,salmon_co-accretion_2022}, but alternative non-impact
possibilities exist, such as those involving spin-orbit interactions between Uranus and primordial satellites \cite{saillenfest_tilting_2022, rogoszinski_tilting_2020} . Whatever the cause of
the tilt, either the orbital plane of the satellites must have re-oriented
at the same time as Uranus, the satellites must have been formed 
by the event itself, or the satellites formed subsequent to the event. The available data do not point to
an obvious favorite among the multiple possibilities. 

Isotopic measurements offer a potentially powerful means of distinguishing between these scenarios. 
{ Decades of laboratory, theoretical, observational, and modeling
work have sought to use measurements of the
ratio of deuterium to hydrogen (hereafter the D/H ratio)
across and beyond the solar system
as a proxy for understanding protoplanetary disk temperature and dynamics, 
inferring the path of water into the inner solar system, and tracing
the inheritance of unprocessed ices from the interstellar medium into the planetary
system \cite{furuya_water_2017, yang_dh_2013, albertsson_chemodynamical_2014}. 
Such measurements have been instrumental in trying to understand such
diverse topics as the delivery of water to the Earth, the accretion
of solid material into Uranus and Neptune, and the provenance of the ices
incorporated into comets \cite{albertsson_chemodynamical_2014, lis_terrestrial_2019, bockelee-morvan_cometary_2015}. }

{ The D/H ratio can be used to trace similar processes in the formation of
the Uranian satellites, and different scenarios for the formation of
the satellites can lead to large differences in predicted D/H ratios.}
The D/H ratio of Uranus has been measured to be 
(4.4 $\pm 0.4)\times10^{-5}$ \cite{feuchtgruber_dh_2013} (about twice 
the solar value), presumably reflecting a mixture of protostellar
H$_2$ carrying the solar D/H ratio, and outer solar system ices with
elevated values, while comets from the outer solar system
have measured D/H values that range from 1.6 to 6.5 $\times 10^{-4}$
\cite{bockelee-morvan_cometary_2015, biver_isotopic_2016, lis_terrestrial_2019, muller_high_2022}, between 4 and 15 times higher than
that of Uranus. { Depending on the source material, the D/H of the Uranian satellites
could plausibly fall anywhere within this range.}

Observations from the Cassini spacecraft and the James Webb Space Telescope (JWST) have shown
that deuterated water can be detected in the reflectance spectrum
of the icy satellites of Saturn \cite{clark_isotopic_2019, hedman_waterice_2024, brown_deuterated_2025} via
a spectroscopic absorption at about 4.14 {\textmu}m caused by
the O-D stretch fundamental (analogous to the well-known 3 {\textmu}m O-H
stretch fundamental). 
Here, we analyze similar JWST 
observations of the major regular satellites of Uranus in an attempt to 
identify the O-D stretch, quantify the D/H ratios on these 
satellites, and constrain the origin of the satellite system.

\begin{figure}
\centering
\includegraphics[width=1.\linewidth]{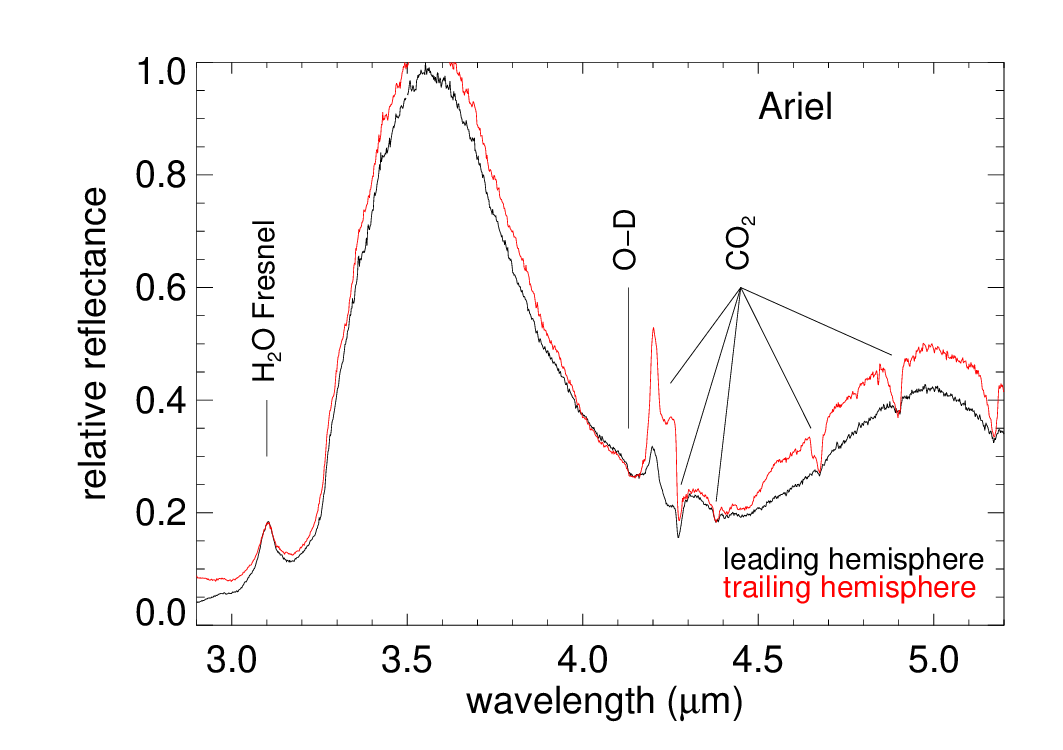}
\caption{JWST near-infrared reflectance spectra of the leading and trailing hemispheres of the
Uranian satellite Ariel \cite{cartwright_revealing_2024} The overall structure of the spectra 
is dominated by the 3 {\textmu}m O-H stretch absorption
from H$_2$O and the broad 4.5 {\textmu}m H$_2$O combination band.
At smaller scales, the 3.1 {\textmu}m Fresnel reflectance peak can
be seen. The region beyond 4 {\textmu}m has multiple features
due to CO$_2$, particularly on the leading hemisphere, including
the 4.26 {\textmu}m $\nu_3$ asymmetric stretch absorption,
a reflectance peak at 4.22 {\textmu}m and multiple weaker absorption
features at longer wavelengths. A small absorption at 4.14 {\textmu}m
is coincident with the expected fundamental O-D stretch feature
of deuterated water, but is also in 
a spectral region potentially containing
absorption due to CO$_2$; the strength of
the 4.14 {\textmu}m feature does not appear to change 
between the leading and trailing 
hemispheres in spite of the large
difference in CO$_2$.}
\label{fig:spectrum}
\end{figure}

\begin{figure}
    \hspace{-.55in}\includegraphics[width=1.25\linewidth]{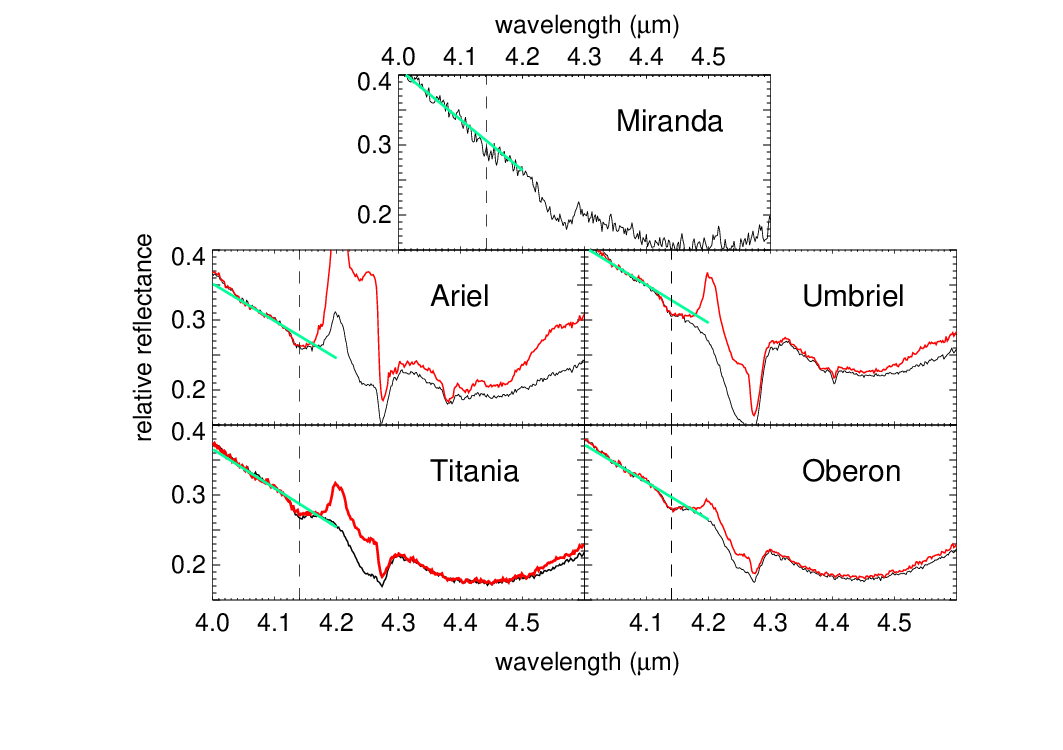}
    \caption{The 4.0-4.6 \textmu m region of the Uranian satellite spectra containing the 4.14 \textmu m O-D absorption
    along with multiple CO$_2$ absorption and reflection features. The leading hemisphere
    spectra are shown in black with the trailing in red. The vertical dashed line shows the
    predicted position of the O-D line for each observation. In all cases, when the continuum
    level is matched between the leading and trailing spectra using the wavelengths just 
    shortward of the 4.14 \textmu m line, the O-D absorption appears essentially identical 
    in spite of substantial variation in the CO$_2$ features across the region. For each spectrum
    a linear fit is performed to the spectrum shortward of the O-D absorption and this estimated continuum
    is projected across the full line as shown in green. Owing to the positive curvature of the spectrum
    at this point, this straight line will underestimate the true area of the O-D absorption, 
    an effect for which we empirically correct using laboratory data.}
    \label{fig:bigfig}
\end{figure}

\section*{Observations and Results}
{JWST observations of the reflectance spectra of
the major satellites of Uranus reveal a spectrum dominated
by crystalline water ice but with additional spectral structure
around the expected 4.14 {\textmu}m O-D absorption feature, including both 
the potential O-D absorption but also possibly overlapping complex CO$_2$
reflection and absorption 
(Figure 1).
To help determine if the absorption seen near 4.14 {\textmu}m could be
due to the O-D stretch,
we first use laboratory experiments
which measured the O-D line position
on similarly crystalline water ice
to find the precise expected location of
the O-D absorption at the temperature of
each satellite (see Supplemental Information).
The location of the absorption seen on each of
the satellites indeed matches the laboratory location 
(see Supplemental Information) suggesting that the
feature that we see is at least partially due to the O-D absorption. 
Next, we assess the
affect of 
the presence of the spectrally adjacent 
CO$_2$ reflectance peaks and broad 
absorptions.
These CO$_2$ features unexpectedly
vary greatly between separate observations of the
leading and trailing hemispheres of each of the largest
four satellites.
By carefully
matching the leading and trailing hemisphere spectra of each
of the individual satellites in the wavelength region just short 
of the expected O-D absorption,
we can see that, in spite of the dramatic variations
in the CO$_2$ abundances between the hemispheres,
the feature at the predicted location
of the O-D absorption is nearly unchanged
(Figure 2 and Supplemental Information). We conclude that these 
absorption features near 4.14 $\mu$m which appear
at the location predicted for O-D and which appear
unchanged in the face of substantial variations in CO$_2$,
are indeed due to the O-D stretch in HDO molecules
within the H$_2$O matrix. 

To measure the strength of the O-D absorption feature we first 
must determine how the spectral continuum would appear in the absence of the absorption.
While we would ideally fit a second-order polynomial to the gently curving
spectral regions adjacent to the absorption to estimate this
continuum, the spectrum
just longward of the O-D absorption is clearly affected by CO$_2$ and cannot be used reliably. 
To minimize contamination from the strong CO$_2$ bands, 
we fit a straight line to the unaffected shorter wavelength
side of the O-D absorption and extend that line through the absorption as
our estimate of the continuum in the absence of the O-D absorption (Figure 2).
We measure the strength of the O-D absorption by fitting a gaussian 
function to the spectrum after it has been divided by this continuum
(Figure 3).
Owing to the positive curvature of this region of the ice spectrum, 
our approximated straight line continuum will underestimate the
true continuum level and thus underestimate the true strength of the absorption.
We calibrate a correction factor and
the uncertainties in the calibration using our laboratory data (see Supplemental Information). 
Table 1 gives the measured and corrected areas of the O-D absorption for
each of the observations.}

\subsection*{The D/H ratio}
{ In order to convert the measured area of the O-D absorption to a D/H
ratio, we use our laboratory data to measure the area of the O-D line 
in water ice doped to have D/H values ranging
from 1.56$\times10^{-4}$ (the VSMOW value) to $7.80\times10^{-4}$ ($5\times$ VSMOW).
We find a linear relationship between the area of the O-D line and the D/H ratio
and apply this calibration and its associated uncertainties (see Supplemental Information)
to our measured and corrected band areas to derive D/H ratios for the 5 Uranian satellites
(Table 1). Figure 4 shows the D/H ratios for the 5 satellites. The remarkable similarity
of the values derived across the full system in spite of considerable variation in 
the CO$_2$ spectrum is again a confirmation that these measurements are unaffected by CO$_2$. 

Most of the measurements are within nearly 1$\sigma$ of the average D/H
value of 2.1$\times 10^{-4}$ of all of the satellites. 
The largest deviation is of the
inner satellite Miranda, which is 2.5$\sigma$ higher than this mean
(or 2.8$\sigma$ above the mean of 2.0$\times 10^{-4}$ if Miranda is excluded).
The Miranda spectrum has the lowest signal-to-noise of the observations, so this deviation
could be a product of that uncertainty, or it could reflect a true 
difference. 
We consider the possibility of an elevated D/H for Miranda below,
but emphasize the need for confirmation of this result.

While the laboratory experiments have been conducted on pure water ice, the Uranian
satellites also contain substantial amounts of dark material,
as evidenced by their albedos ranging from 0.19 to 0.32 \cite{buratti_surface_1990}. No clear spectral signatures
of this dark material have been detected \cite{cartwright_revealing_2024}, but
mixing of any spectrally plausible 
dark material with water ice will affect the O-D absorption and its surrounding spectral regions
equally, leading to no discernible effect on the continuum-corrected band area.}
\begin{table*}
\centering
\caption{Absorption depths and D/H ratio on Uranian satellites}
\begin{tabular}{lccccc}
satellite & hemisphere & 4.14 {\textmu}m area& corrected area & D/H ratio &albedo$^2$  \\
 & & ($\times 10^{-3}$ \textmu m)$^1$ &($\times 10^{-3}$ \textmu m )$^1$ & ($\times 10^{-4}$) \\
\midrule
Miranda &leading    &$ 1.9\pm0.2$ &$ 2.3\pm0.2$ &$ 3.1\pm0.4$ &  0.33\\
Ariel   & leading &$ 1.4\pm0.1$ &$ 1.6\pm0.1$ &$ 2.2\pm0.3$  & 0.34\\
        & trailing&$ 1.3\pm0.3$ &$ 1.6\pm0.3$ &$ 2.1\pm0.5$  \\
Umbriel & leading &$ 1.3\pm0.2$ &$ 1.4\pm0.2$ &$ 2.0\pm0.4$  & 0.19\\
        & trailing&$ 1.1\pm0.2$ &$ 1.2\pm0.2$ &$ 1.7\pm0.4$  \\
Titania & leading &$ 1.1\pm0.3$ &$ 1.3\pm0.3$ &$ 1.7\pm0.5$  & 0.27\\
        & trailing&$ 1.6\pm0.1$ &$ 1.9\pm0.1$ &$ 2.5\pm0.3$  \\
Oberon  & leading &$ 1.4\pm0.1$ &$ 1.6\pm0.1$ &$ 2.1\pm0.3$ &  0.19\\
        & trailing&$ 1.4\pm0.1$ &$ 1.7\pm0.1$ &$ 2.3\pm0.3$  \\
\bottomrule
\end{tabular}

\addtabletext{\hspace{1.3in} $^1$The continuum-normalized band area, sometimes called the equivalent width, has units of \textmu m. \\
\hspace{-2.3in} $^2$ Average clear filter normal reflectance \cite{buratti_comparative_1991}.}
\end{table*}
\begin{figure}
    \centering
    \includegraphics[width=1\linewidth]{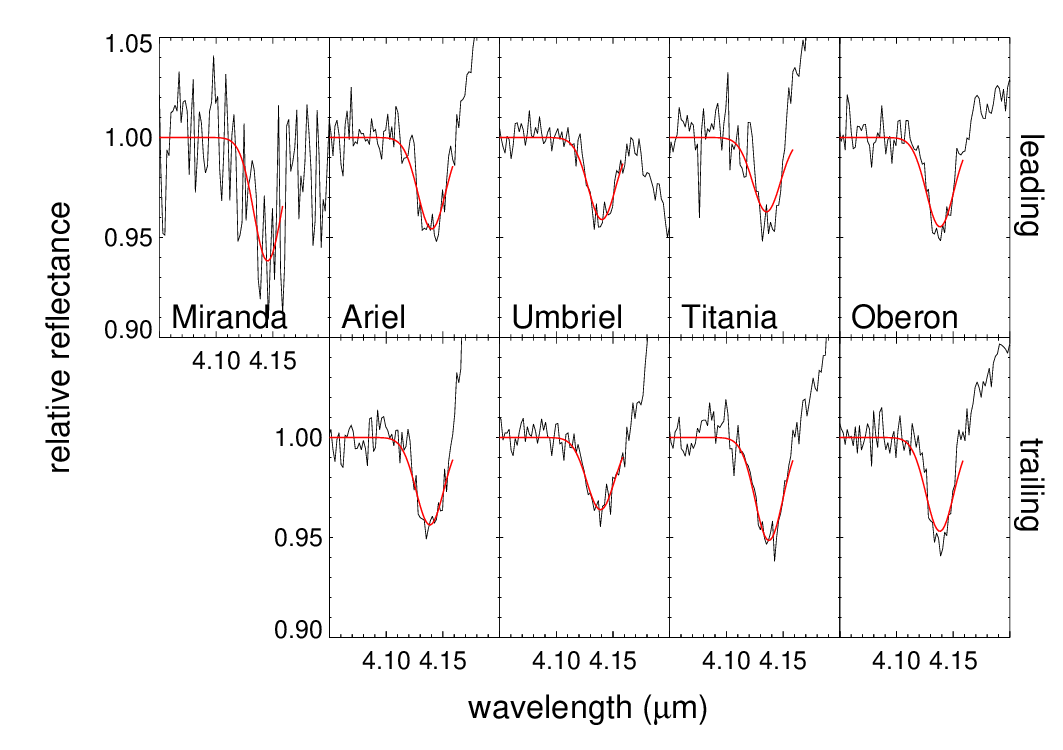}
    \caption{Continuum-divided spectra of the Uranian satellites highlighting the 
    4.14 \textmu m absorption region. The red lines shows the gaussian fit
    to the absorption features and illustrate the spectral region over which the
    function was fit. The region of the spectrum just longward of the 4.14 $\mu$m
    absorption can be highly affected by CO$_2$ absorption or reflection and so is 
    carefully excluded from the fit.}
    \label{fig:placeholder}
\end{figure}
\begin{figure}
    \hspace{-.1in}\includegraphics[width=1.\linewidth]{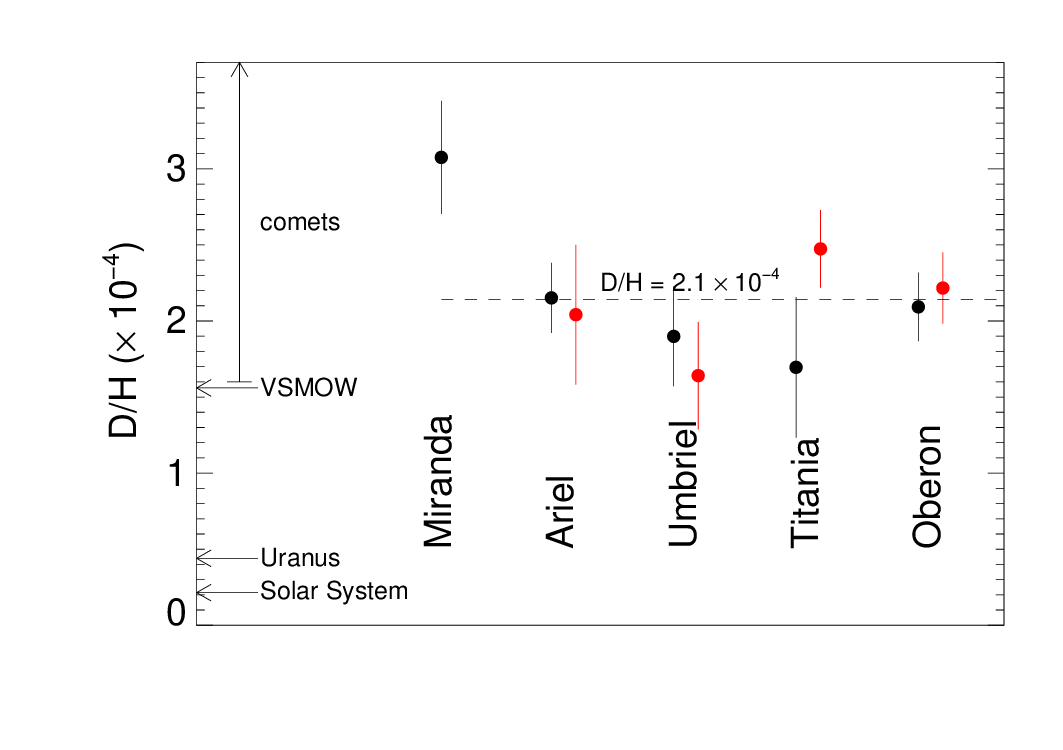}
    \caption{The measured D/H ratio for the satellites of Uranus. Leading hemisphere
    observations are show in black, while trailing are in red. The average of
    all satellite is
    $(2.1\pm0.2)\times10^{-4}$ or $(2.0\pm0.2)\times10^{-4}$ if Miranda is 
    excluded from the average.}
    \label{fig:placeholder}
\end{figure}

\section*{Implications for the formation of the satellites}
{ The (2.1$\pm 0.2)\times 10^{-4}$ average D/H ratio inferred from these data is almost 5 times} higher than that measured in the atmosphere of
Uranus \cite{feuchtgruber_dh_2013} and comparable
to values measured in comets
\cite{biver_isotopic_2016, muller_high_2022}. Moreover, 
{ with the possible exception of the elevated value for Miranda},
no evidence for variation among the satellites is seen. We use these
observations to evaluate the history of the Uranian satellites.

\subsection*{Contamination}
{ One source of water 
with an elevated D/H ratio could be
external impactors contaminating the
surfaces of the satellites.
By far the largest source of such material is
dust generated from
the long-term collisional erosion of the
irregular satellites, some of which spirals inward
to intercept the regular satellites \cite{bottke_irregular_2010}.
While the current flux of dust is trivial,
early in the history of the solar system the irregular
satellite system would have been more extensive,
and satellite-satellite collisions would have
led to  substantial dust clouds \cite{bottke_irregular_2010, kennedy_collisional_2011}.
It is difficult, however, to provide even order-of-magnitude estimates of the
total dust delivery to the regular satellites during this time
owing to the complications of dust collisions, magnetospheric 
interactions, and chaotic dynamics \cite{kennedy_collisional_2011, tamayo_chaotic_2013}.

While the total mass delivered to the
regular satellites is unknown, simulations
suggest that $\sim$3 times more mass is
delivered to the leading, rather than trailing,
hemispheres, and that the dust delivery is a strong
function of distance from Uranus, with the outermost satellites
receiving the most (potentially by a large margin)
\cite{tamayo_chaotic_2013}. These suggestions
are consistent with observations of
the leading/trailing albedo and color
differences of the satellites \cite{buratti_comparative_1991, cartwright_red_2018},
and suggest a light but not complete covering of the surface by irregular 
satellite debris.
Notably, however, we see no evidence for a
systematic difference in D/H ratio 
across the hemispheres nor any evidence for
major differences across the outer four
satellites, suggesting that contamination of the water
ice has little effect on the measured D/H ratio.

Further evidence for the insensitivity of
these measurements to external contamination
is the consistency  of the measured D/H
ratios  across the outer four satellites
in spite of the $\sim$50\% differences in albedo (Table 1). These
albedo differences are mostly controlled by the differences in
exposure of 
(presumably native) water ice from impacts and tectonics,
but this exposure does not appear to change the measured D/H ratio.
Some of the albedo differences relate to the differences in surface ages
between the satellites; while the three outermost satellites
appear to have ancient surfaces, Miranda and Ariel have regions
significantly younger \cite{zahnle_cratering_2003, bottke_bombardment_2024}.
As the flux of irregular satellite dust decreases rapidly with time, 
such large differences in surface age would again lead to significant differences
in the amount of accumulated irregular satellite material,
yet no effect on the D/H ratio is seen.
Based on all of these arguments, we conclude that we are measuring the D/H ratio of the native
water ice on each satellite.
}

Related to contamination is the possibility of
post-formation D/H fractionation.
Such a possibility has been considered for
the satellites of Jupiter \cite{bierson_explaining_2020},
and, even considering the substantial water loss possible at Europa,
an increase in the D/H ratio of only $\sim$10\% 
is predicted, with smaller enrichments
as a function of distance from Jupiter. Similar modeling trying to explain
the previously-believed high D/H ratio of 
Saturn's satellites Phoebe 
(now known to be in error \cite{brown_deuterated_2025}) found that even an extreme 50\% post-formation water loss resulted in less than a 50\% increase in the D/H ratio \cite{clark_isotopic_2019}. 
Given the small predicted magnitude
of these effects, we regard fractionation
as unimportant at these scales.

\subsection*{Formation in a post-impact disk}
The tilt of Uranus and the existence of a satellite system could both be plausibly
explained by a single giant impact \cite{safronov_sizes_1966, slattery_giant_1992}. Models of the disk generated from the giant impact
suggest that material from the atmosphere and envelope  of Uranus mixes with the material from the impactor in a
viscously spreading fully vapor disk; the vapor then condenses to form solids 
out of which the satellites accumulate \cite{ida_uranian_2020}.
{ Detailed models of such a collision have had a difficult time
simultaneously reproducing key properties of the Uranian system including
the mass, angular momentum, and composition, except in the case
of a nearly pure rocky impactor \cite{reinhardt_bifurcation_2020, kegerreis_consequences_2018, woo_did_2022}.
For the extreme case of a 100\% rocky impactor, all of the water in the
satellites would have to come from the envelope of Uranus.
The atmosphere and envelope of Uranus are often assumed to have be completely mixed
\cite{feuchtgruber_dh_2013}, implying that the envelope shares the
same low D/H ratio as the measured atmospheric H$_2$. The initial
disk would thus have the Uranian D/H ratio of $4.4\times 10^{-5}$.
Cooling of the disk would allow deuterium enrichment of the water vapor
at the expense of any H$_2$, but even in an extreme case of a very water vapor
poor disk, the maximum enrichment before condensation is only 
about 30\% \cite{mousis_modeling_2004}. Even with this enrichment,
the water ice would still have a D/H ratio nearly four times lower than
that measured here. Creation of the satellites from a vapor disk created by a 100\% rocky impactor
is clearly ruled out.

If the assumption of a 100\% rocky impactor is
relaxed, water from the impactor -- potentially with a higher D/H value -- will be 
mixed with the water from Uranus. Simulations have found, however, that even 
a relatively water-poor object with only a 27\% fraction of water ice (similar to
those inferred for Triton and Pluto) places far too little rock into the
disk to yield the current densities of the Uranian moons \cite{woo_did_2022}. Bodies with even less
water have yet to be studied, so perhaps a delicate balance can be achieved between
increasing the rock fraction to inject more rock into the disk and maintaining
sufficient water in the impactor to raise the D/H ratio of the satellites.
Assuming that the impactor had a D/H ratio of $\sim 4\times 10^{-4}$, a
mid-range cometary value, approximately half of the water ice
in the disk would have to come from the impactor to reach our measured D/H
ratio. It is clear that the D/H ratio measured here places a strong constraint
that could plausibly rule out any such models.}

\subsection*{Formation from a ring of tidally disrupted outer solar system bodies}
An alternative scenario for the formation of the satellites of Uranus is 
the coalescence of material from a Roche-limit-interior ring system and subsequent outward
tidal evolution to where
they appear today \cite{charnoz_recent_2010, crida_formation_2012}.
This hypothesis was originally developed to explain the small moons just
exterior to the rings of Saturn  but then generalized to potentially
explain satellite systems throughout the solar system. 
Satellites formed in this way naturally increase in mass as a function
of distance and models can approximately reproduce
the mass of the Uranian satellites \cite{crida_formation_2012}. The lower density of Miranda,
in this case, could be an outcome of accretion in a mixed rock, ice ring system. 
The cores of the outer four satellites would form
from the rocky disk, which has a closer Roche limit, and these cores would accrete ice and
move outward \cite{blanc_understanding_2025}. Miranda would be the last satellite accreted, after much of the rock had already been depleted. 

The most likely
source of the early ring would be tidal disruption 
of passing cometary or Kuiper belt bodies \cite{dones_recent_1991, hyodo_ring_2017}. 
Estimates of the population of large icy bodies passing Uranus in the early solar system
are sufficient to have led to several encounters large enough to build the Uranian
satellite system \cite{hyodo_ring_2017}.
If
this disruption occurred after the tilt of Uranus,
the ring would naturally lie in the equatorial 
plane of the planet, separating the Uranus-tilting
event from the creation of the satellites. Tidal evolution 
to the orbital distance of Oberon has long been thought to be 
difficult, but recent analysis suggest that -- as at
Saturn -- tidal migration perhaps proceeds much more quickly than 
previously believed \cite{nimmo_strong_2023, jacobson_orbits_2025}.

In this scenario, the material that forms the
satellites never mixes with material from Uranus,
so the original D/H ratio of the progenitor is 
preserved. The D/H ratio of the Uranian satellites is on the low
end of values found in comets, consistent with a progenitor from
this source population, though measuring the D/H ratio in water ice on some of the larger dwarf
planets in the Kuiper belt is an important step towards understanding 
typical deuterium abundances in the most massive members of this population.

{ One potential difficulty with this scenario is the possibly elevated D/H ratio
of Miranda. All satellites should share the D/H ratio of the progenitor. 
Confirmation of an elevated D/H ratio for Miranda would be difficult to
reconcile with this
formation mechanism.}

\subsection*{Formation from re-accreted early satellites}
A final model we consider 
suggests that the satellites and their large tilt to the plane of the solar
system could be explained by the combination of a giant impact Uranus tilting event, followed
by the collisional disruption of a pre-existing satellite system, leading to the re-accumulation of
the satellites in the tilted equatorial plane of Uranus \cite{morbidelli_explaining_2012}. 
The reorientation of the re-accumulated satellites would
require a compact massive impact-generated disk interior to the orbits of
the satellites to boost the effective J$_2$ value
of Uranus. In such a scenario there is no exchange with the interior
of Uranus and the
original D/H ratio of the primordial satellites would be retained.

{ At Saturn, the D/H ratio of the satellites is substantially higher than the
solar-like value seen at the planet itself \cite{waite_jr_liquid_2009,brown_deuterated_2025}, presumably reflecting a 
formation from the accretion of solid ices from the proto-solar nebula to construct
the satellites, rather than from condensation of material in a proto-Saturnian
nebula. A similar process could hold at Uranus, suggesting that a pre-existing 
satellite system would have an elevated D/H ratio compared to Uranus.

The inner satellite Miranda poses an interesting challenge for
this model. Its substantially lower density is difficult to
explain if it forms from the same reaccumulated satellites,
but it could possibly have formed from the compact impact-generated
disk itself \cite{hesselbrock_three_2019, rufu_coaccretion_2022, salmon_co-accretion_2022}.
In this case the D/H ratio of Miranda could be substantially
different from the other satellites. Indeed, 
if the impactor is moderately icy (which is allowed
in this scenario, as the disk does not need to provide
the high rock fraction for the outer satellites) and much
of the water from the disk is due to the impactor, an elevated
D/H ratio for Miranda would suggest a typical cometary D/H 
for the impactor.
 
Detailed simulations have not tracked the source of water
for the inner disk that may coalesce into Miranda, nor have they
successfully reproduced all of the neccessary condition for this 
scenario \cite{rufu_coaccretion_2022, salmon_co-accretion_2022},
but the decoupling of the D/H ratio of Miranda from
that of the other satellites is a unique feature of this model
that is worth further exploration.}

Water isotopes provide a powerful probe into the source and history of the material
that formed the satellites of Uranus. { The elevated D/H ratio of the Uranian satellites
strongly precludes formation from material that has substantially mixed with the low D/H ratio
material of Uranus through either dilution or equilibration 
and instead suggests that the original material 
never interacted with Uranus itself. The inner satellite
Miranda holds particularly powerful clues to the formation of the satellites.
A D/H ratio at Miranda that differs from that at the other satellites is
difficult to explain if the satellites formed from the disruption and re-accretion
of a tidally captured outer solar system body. If, alternatively, the outer four
satellites are re-accreted fragments of a pre-existing satellite system
and Miranda coalesces out of an impact-derived vapor disk, its D/H ratio
could deviate from the others. The measurement that Miranda has a D/H
ratio 2.5$\sigma$ higher than the average of the other satellites is
suggestive, but not conclusive, of this scenario. Higher signal-to-noise spectra
will need to be obtained to confirm this finding.

An additional set of observations could also help to constrain 
the formation of these satellites. While cometary values show
a wide range of D/H, it is not known what the average value of D/H in the
water of the outer disk would have been. The large bodies that would have
impacted Uranus would presumably have D/H ratios typical of this overall
average rather than the range of individual comets. Measurement of the
D/H ratio in water ice on some of the larger dwarf planets would help to 
determine this average value.}

\acknow{We gratefully acknowledge conversations with Sebastien Charnoz and Francis Nimmo
about this work and insightful comments from the reviewers on the manuscript. This work is based on observations made with the NASA/ESA/CSA James Webb Space Telescope. The data were obtained from the Mikulski Archive for Space Telescopes (MAST) at the Space Telescope Science Institute, which is operated by the Association of Universities for Research in Astronomy, Inc., under NASA contract NAS 5-03127 for JWST. These observations are associated with program \#4645, with archival observations
from program \#1786. Support for program \#4645 was provided by NASA through a grant from the Space Telescope Science Institute, which is operated by the Association of Universities for Research in Astronomy, Inc., under NASA contract NAS 5-03127.}

\showacknow{} 

\clearpage

\section*{Supporting information}

\subsection*{Observations}
Spectra of the satellites of
Uranus were obtained using the JWST NIRSpec
near-infrared spectrometer from JWST programs \#4645
and \#1786. The data were taken using the medium spectral
resolution G395M grating, which covers the wavelength
range from 2.9 to 5.2 $\mu$m, and the integral field unit (IFU).
Each target moon was dithered over four separate positions
on the IFU to help remove systematic artifacts.
The observational details for the 
four outer satellites are described elsewhere \cite{cartwright_revealing_2024}. 
Miranda was observed between  	Feb 7, 2025 16:19:19-18:11:08 UT, corresponding to 124$^\circ$ to 143$^\circ$ sub-observer longitude. The observations were taken using both the G235M and G395 gratings, with 24 and 20 groups per dither over 4 dithers, corresponding to 1459 and 1225 seconds of exposure time respectively.
The spectra were extracted from the observations using
a modified version of the standard JWST pipeline \cite{brown_state_2023, brown_JWST_2025}.
{\ The IFU provides the spectrum of both the satellites and the blank sky 
surrounding it, allowing accurate subtraction of any scattered light from
nearby Uranus.}
To obtain relative reflectivities, the extracted spectra
were divided by the spectrum of the star P330-E, a spectral analog
to the sun. The spectrum of Miranda, median smoothed by 10 pixels to remove
bad pixels, is shown in Figure~5.
\begin{figure}[h]
\vspace{-25pt}
\centering
\includegraphics[width=0.5\textwidth]{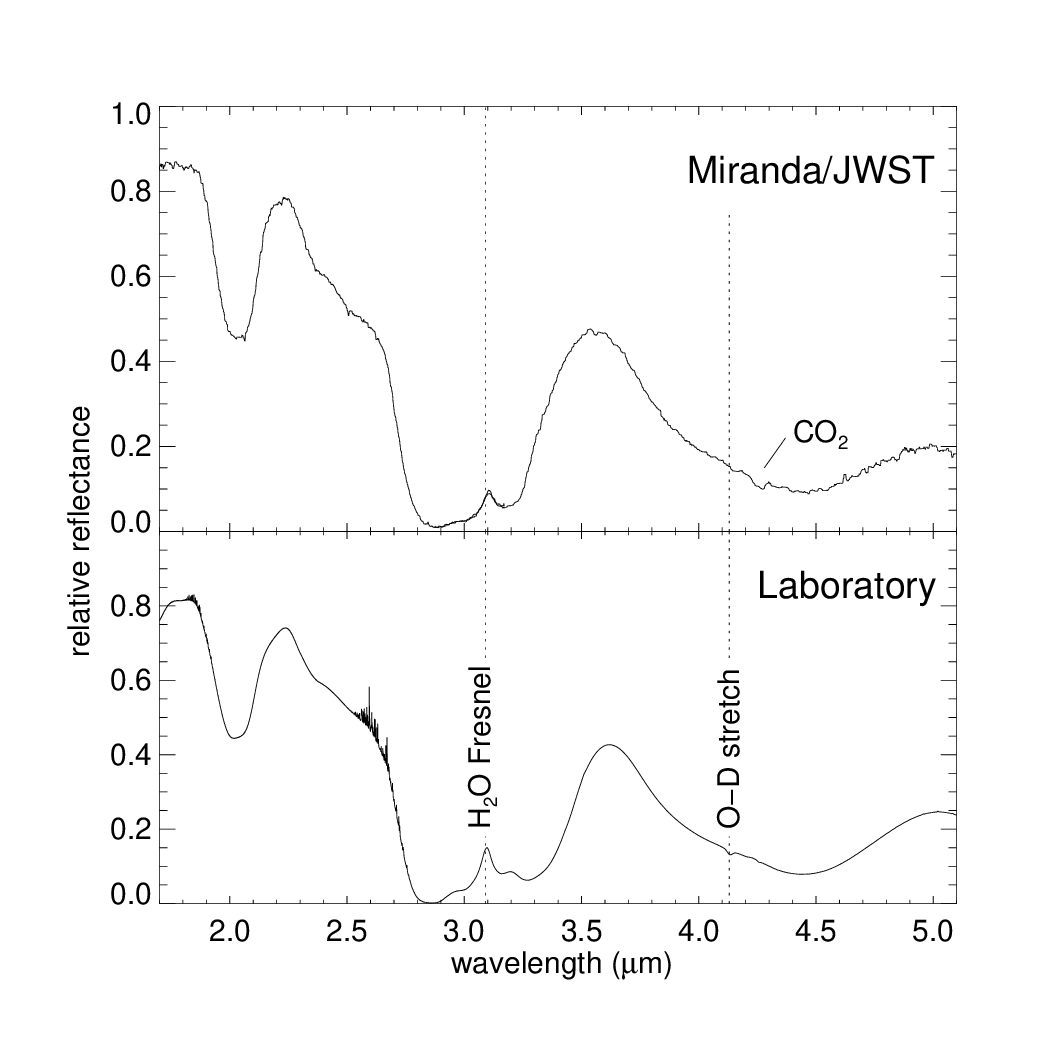}
\vspace{-30pt}
\caption{Top: The JWST spectrum of Miranda, showing the H$_2$O Fresnel peak, the location of the O-D stretch
feature (barely visible at this scale), and the CO$_2$ absorption. Bottom: A laboratory reflectance spectrum of H$_2$O doped
with HDO. No CO$_2$ is present in this spectrum. The difference in
temperature between the sample in the laboratory spectrum and Miranda
can be seen in the shift of Miranda's Fresnel peak to slightly longer
wavelengths, the large shift of the $\sim$3.6 $\mu$m band to
shorter wavelengths, and the shift of the O-D feature
to longer wavelengths. The features near 1.85 and 2.6 $\mu$m in the laboratory
spectrum are residual uncorrected water vapor bands. Note that the strength of
the Fresnel peak is highly dependent on the surface texture of the ice, so it
varies greatly even for purely crystalline water ice.}
\vspace{-10pt}
\end{figure}

\subsection*{Laboratory spectroscopy}
{ To understand the effect of deuterium on the spectrum of water
ice, 
we performed two suites of spectral measurements of 
water ice doped with varying amounts of deuterated water.
In the first set of experiments, we made 60 separate deuterated water 
samples in a nitrogen-purged glove box by mixing deionized water
with calibrated volumes of pure 
D$_2$O to reach a desired final D/H ratio. The
sealed liquid sample was allowed to equilibrate
for 120 minutes at room temperature and then transferred to
a commercial mechanical nebulizer. The nebulizer produces fine droplets
by the use of a vibrating membrane, a process which should preserve the
isotopic composition of the water. The nebulized water was sprayed onto 
the} 
liquid nitrogen cooled sample
holder of a low temperature reaction chamber (CHC-CHA-4 from
Harrick Scientific). The reaction chamber was then sealed with a
three-port dome (equipped with two ZnSe and one quartz window),
removed from the glovebox, and mounted onto
a Praying Mantis diffuse reflectance accessory, also from Harrick
Scientific, within the sample compartment of a Nicolet iS50 FT-IR
spectrometer. The chamber was then evacuated to rough vacuum (0.03
bar). The Praying Mantis is equipped with two \ang{90} off-axis ellipsoidal mirrors one of which focuses the incident IR beam onto the sample while the other collects the diffuse reflection from the sample across a large solid angle (accounting for $\sim$20 \% of reflected light).  

{ Use of the nebulizer allowed the production of water ice with a spectrum closely
resembling those of the Uranian satellites, including a prominent 3.1 $\mu$m
emission peak indicative of crystalline water ice and a downward sloping
spectrum from 2.2 to 2.8 $\mu$m associated with diffraction from sub-micron 
particles \cite{clark_isotopic_2019}.
An example lab
spectrum compared to the spectrum of Miranda is shown in Figure 5.}

While a thermocouple in the sample gave a measurement of the temperature at some depth in the sample, the thermal
gradient in the sample ensured that the temperature of the surface -- exposed to the thermal
radiation of the chamber walls -- was higher. The location of the peak in
reflectance between the 3.0 $\mu$m and 4.5 $\mu$m band is known to be temperature
dependent \cite{filacchione_saturns_2016}, so we used this band to constrain our temperatures, which we
found to range from 110 to 180 K, higher than the
$\sim$85 K temperatures typical of the Uranian satellites
\cite{hanel_infrared_1986, detre_herschel-pacs_2020}.

{ To ensure that our measurements remain valid at the temperature of the Uranian
satellites, we performed a second set of 12 experiments in an ultra-high vacuum 
chamber with a closed-cycle helium cooled cold finger. For these experiments,
the deuterated water was produced identically to the first process, then
the nebulized water was sprayed directly onto a liquid nitrogen-cooled copper sample
holder which was then placed onto the cold finger in the chamber. The chamber was pumped to
a pressure of 10$^{-7}$ torr and the cold finger cooled to cryogenic temperatures
simultaneously. The same FTIR was used to observe the sample, but a co-addition
of 9450 scans was required to achieve adequate signal-to-noise through this apparatus.
Based on comparison to the position of the 3.1 $\mu$m water Fresnel feature 
(see below), these experiments obtained temperatures as cold as those on the
Uranian satellites. 

The position of the O-D absorption band
was preliminarily estimated by fitting a first-order
polynomial continuum
to the region between 4.0 and 4.1 $\mu$m and dividing the
spectrum by that continuum. 
We then fit a gaussian function to the continuum-divided region at 4.14 $\mu$m
and determined the approximate
center of the absorption feature.
Because of the wavelength dependence of the absorption line (see below),
we refined our fit by more carefully
fitting a second-order polynomial to the 
data from 0.105 $\mu$m shortward of the measured center of the O-D line to 0.75 $\mu$m
longward of the line while excluding the region within 0.03 $\mu$m of the center.
We divide the original spectrum by this new continuum and fit a gaussian function
to the resulting line using a non-linear least-squares algorithm. We record
the gaussian width, height,  and center for each spectrum. The band area 
(called equivalent width in the astronomical literature), in
units of $\mu$m for these continuum normalized spectra, is calculated from the gaussian
fit. }

\subsection*{JWST band position and area} 
{ To fit the position and area of the O-D absorption in the JWST data, we use
only the short wavelength region of the adjacent spectra for construction of
the continuum, as the longer wavelength region is clearly affected by CO$_2$ in
many cases. We construct the continuum by fitting a straight line to the region
from 4.0 to 4.1 $\mu$m and extending that straight line through the O-D feature.
We divide the spectrum by this estimated continuum and fit a gaussian to the remaining
absorption between the wavelengths of 4.04 and 4.16 $\mu$m. Note that this wavelength
excludes the longest wavelength of the absorption line which might be affected by
CO$_2$. We fix the width of the gaussian to be the 0.0122 $\mu$ mean width of the
lines measured in the laboratory spectra. 

We also fit the line position of the temperature-dependent H$_2$O Fresnel peak
at approximately 3.1 $\mu$m. The Fresnel peak shape is poorly
approximated by a gaussian function, so we fit a second order
polynomial to the spectrum between 3.09 and 3.12 $\mu$m and use the peak as the
line position. An identical procedure is performed to the laboratory data.}

\subsection*{Temperature-dependent line position}
{ The combined experimental dataset finds
that the wavelength of the O-D stretch feature is inversely proportional
to temperature and can range between 4.121 to 4.142 $\mu$m over the temperature
range in our samples. 
We find a very tight linear correlation
between the location of the 3.1 $\mu$m Fresnel peak, which is also temperature dependent \cite{mastrapa_optical_2009},
and the location of the 4.14 $\mu$m band (Figure 6).
While the peak near 3.6 $\mu$m has been used for temperature
measurements on planetary surfaces before, it is also known that this peak can be 
shifted with non-water ice contaminants \cite{fischer_spatially_2017}. The Fresnel peak is only caused
by water ice, so it is a more reliable temperature proxy for planetary surfaces.
Note that previous laboratory experiments suggested that the location
of the 4.14 $\mu$m peak was line strength dependent \cite{clark_isotopic_2019}. 
We find no evidence for such an effect and suspect that those experiments
were actually being affected by temperature differences, rather than abundance
differences.
\begin{figure}
\centering
\includegraphics[width=0.5\textwidth]{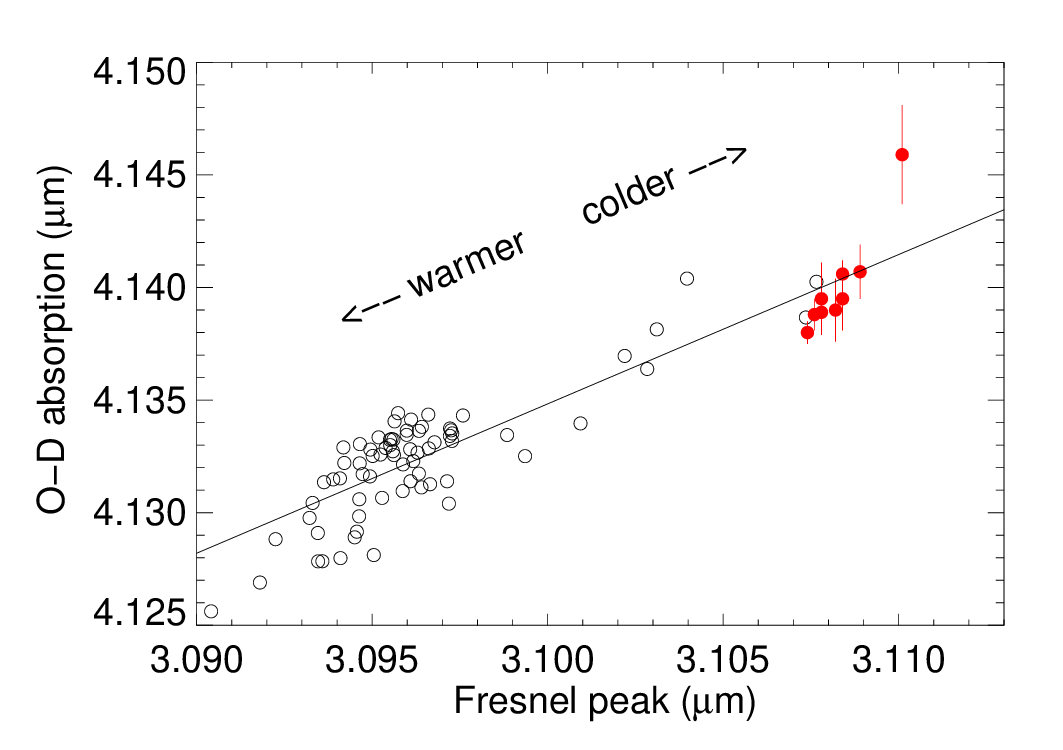}
\vspace{-20pt}
\caption{Measurements of the peak position of the
H$_2$O Fresnel peak and the O-D stretch absorption
in laboratory spectra (black). Both line positions are
temperature dependent, and the laboratory data ranged in
temperature from 85 to 180 K. The line is a linear 
least-squares fit to these data. The red points show the same
measurements for the Uranian satellites, which are 
at temperatures of $\sim$85K \cite{hanel_infrared_1986, detre_herschel-pacs_2020}. These 
measured points fit well,
giving confidence that the 4.14 $\mu$m features observed
are due to deuterated water. Miranda
has the longest wavelength Fresnel peak and O-D absorption and is a 
2$\sigma$ outlier.}
\vspace{-9pt}
\end{figure}
\begin{figure}
\vspace{-15pt}
    \centering
    \includegraphics[width=1\linewidth]{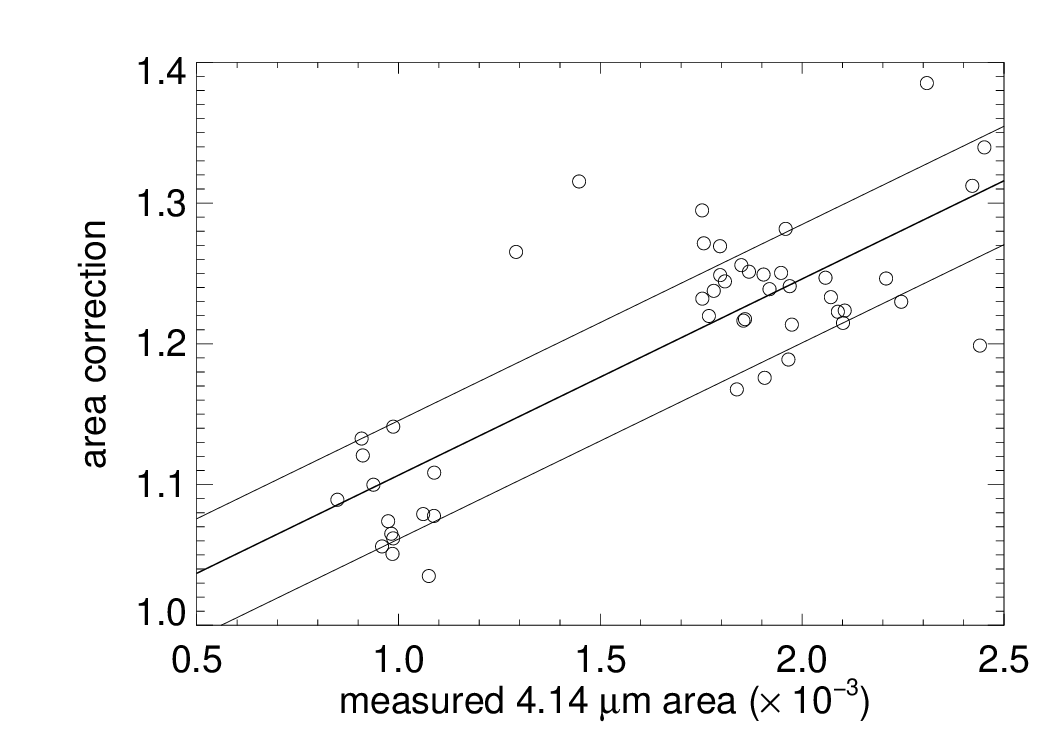}
    \vspace{-20pt}
    \caption{The fraction correction factor found in our laboratory measurements
    of the band area when we use the single-sided short wavelength region 
    of the spectrum to construct a linear continuum to when we use the full
    region and fit a second order continuum. The data reveal a linear
    function between the measured area and the correction factor, and we derive
    uncertainties in this correction factor finding a parallel line
    that is a lower limit
    below 84\% of the data and similar line that is an upper limit above
    84\% of the data. }
    \label{fig:placeholder}
    \vspace{-10pt}
\end{figure}

Figure 6 also shows the location of the Fresnel peak measured on the Uranian satellites
as well as the center of the absorption near 4.14 $\mu$m. The location
of the absorption is consistent with the predicted location based on the temperature-dependent location of the Fresnel peak, giving confidence that these absorption are indeed the expected O-D
stretch feature. Miranda, with a lower signal-to-noise than the other spectra, is a 
2$\sigma$ outlier.}

\subsection*{Continuum uncertainty}
{ The exact manner in which the continuum is estimated can systematically affect the
measured band area. In particular, for our data we fit a linear continuum to the short
wavelength side of the absorption line. For the positively curved water ice
continuum at this wavelength, such a linear fit will underestimate the continuum
and thus underestimate the true band area. To correct for and understand the uncertainties
in using this continuum, we use our laboratory data
to compare band areas measured with a one-sided linear continuum fit to those
measured with a full second-order continuum fit. Figure 7 shows the correction
factor derived for each experiment. The correction appears linear as a function of
the measured band area.
Uncertainties in the continuum calibration are estimated from the bounding lines
which encompass 14 and 86\% of the data, as shown in Figure 7.}
\begin{figure}   
    \centering
    \includegraphics[width=1\linewidth]{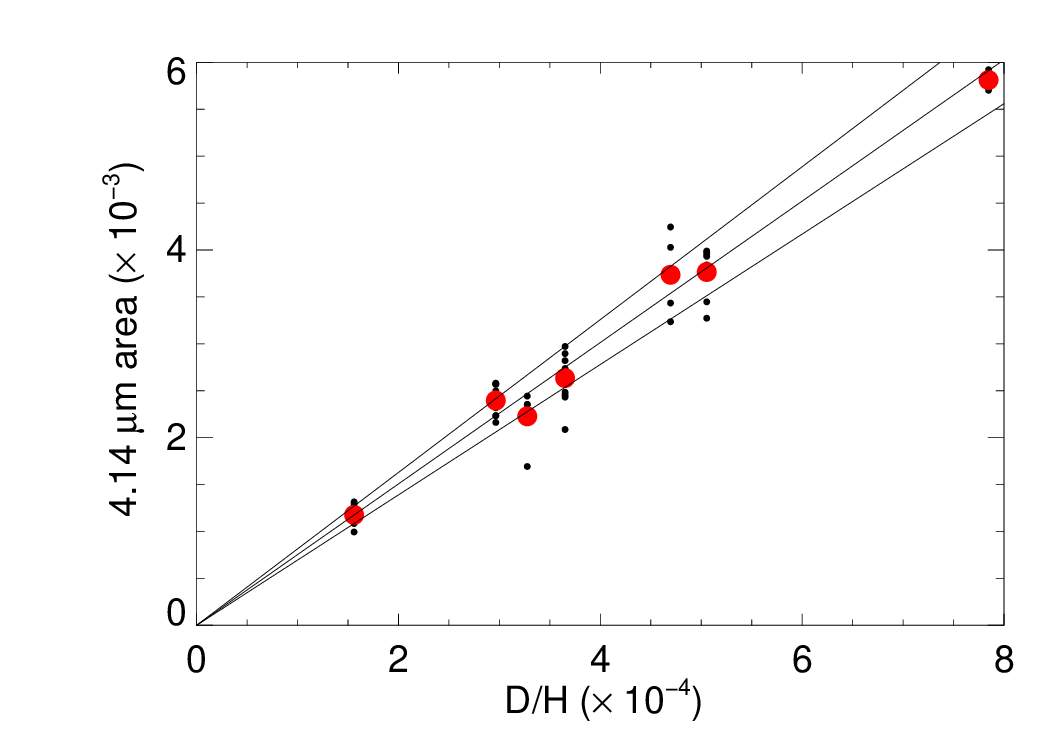}
    \vspace{-20pt}
    \caption{The measured band area (in units of \textmu m) as a function of
    D/H ratio in our laboratory samples. The average of the data at each 
    concentration is shown as a large red point, while the individual experiments are
    shown as small black points. A line, constrained to go through the
    origin, fits the data with a slope of 0.133. The upper and lower uncertainties
    are found from the lines which are above and below 84\% of the data
    respectively. The overall uncertainty to the calibration is 7.5\%. }
    \label{fig:placeholder}
    \vspace{-10pt}
\end{figure}

\subsection*{D/H ratio calibration}
{ 
We use our laboratory data to calibrate the D/H ratio as a function of band
area. The D/H ratio appears to be a linear function of area, so
we perform a linear
fit to the data (with the constraint of zero band area for zero concentration). The
uncertainties in the calibration are obtained from lower and upper boundaries to the
fit which lie above 86\% or below 86\% of the data. The fits and uncertainties can
be seen in Figure 8. Numerically, the calibration is given as $r=0.133 a$, where $a$ is the 
area (equivalent width) of the O-D absorption in $\mu$m and r is the derived D/H ratio. The
uncertainty is $0.010 a$, or approximately 7.5\%.

Note that a previously proposed calibration \cite{clark_isotopic_2019} involved taking a ratio
of the depth of the 4.14 and 2 $\mu$m lines. This proposal was based on a theoretical
spectral model
and three laboratory points and was
expected to correct for variations
due to grain size which will affect
band area. Our series of 72 experimental points shows no evidence for
the validity of this proposed calibration and instead reveals this much simpler linear
correlation with the 4.14 $\mu$m line area. Because the 4.14 $\mu$m O-D
absorption lies within the broad $\sim$4.5 \textmu combination band
of H$_2$O, the measurement of the
continuum-normalized band area 
is itself a ratio of two unsaturated
band areas, effectively 
correcting for grain-size effects.}


\clearpage




\end{document}